\documentclass[aps, prl, floatfix,
 reprint,
]{revtex4-2}

\usepackage[colorlinks,linkcolor=blue,citecolor=blue]{hyperref}

\usepackage{amsmath,amsthm,verbatim,amssymb,amsfonts,amscd, graphicx}
\theoremstyle{plain}

\usepackage{natbib}

\newcommand{\V}{\text{Var}}
\begin{document}

\title{Universal Thermodynamic Uncertainty Relation in Non-Equilibrium Dynamics}% Force line breaks with \\

\author{Liu Ziyin$^1$}
\author{Masahito Ueda$^{1,2,3}$}%
\affiliation{%
$^1$Department of Physics, The University of Tokyo, 7-3-1 Hongo, Bunkyo-ku, Tokyo 113-0033, Japan\\
$^2$RIKEN Center for Emergent Matter Science (CEMS), Wako, Saitama 351-0198, Japan \\
$^3$Institute for Physics of Intelligence, The University of Tokyo, 7-3-1 Hongo, Bunkyo-ku, Tokyo 113-0033, Japan
}%

\date{\today}% It is always \today, today,

\begin{abstract}
We derive a universal thermodynamic uncertainty relation (TUR) that applies to an arbitrary observable in a general Markovian system. The generality of our result allows us to make two findings: (1) for an arbitrary out-of-equilibrium system, both the entropy production and the \textit{degree of non-stationarity} are required to tightly bound the strength of a thermodynamic current; (2) by removing the antisymmetric constraint on observables, the TUR in physics and a fundamental inequality in theoretical finance can be unified in a single framework.

\end{abstract}

\maketitle

\textit{Introduction}. Nature is rife with nonequilibrium phenomena such as expansion of the universe \cite{Kawasaki_1999, berges2015nonequilibrium, berges2004prethermalization}, relaxation dynamics of condensed matter systems \cite{polkovnikov2011colloquium, kamenev2011field}, interactions in ecological systems \cite{rohde2006nonequilibrium} and biological processes in living creatures \cite{fang2020nonequilibrium, wang2015landscape}. Nonequilibrium phenomena are not exclusive to natural sciences. The financial market is also a far-from-equilibrium system, with its complex dynamics arising from interactions among a large number of investors in the market \cite{lux2009applications, samanidou2007agent, dinis2020phase}. The learning dynamics of deep neural networks is also regarded as an important nonequilibrium phenomenon \cite{saxe2013exact, baity2018comparing, zhiyi2021distributional, liu2021noise}. With such a wide range of applications, one naturally wonders what common features, if any, could be shared by all of these heterogeneous systems. Only after knowing what is shared across all nonequilibrium phenomena can we understand what is unique to each individual field and hope to unify nonequilibrium phenomena in different fields of natural and social sciences. This work studies the common features shared among various nonequilibrium problems and establishes a universal thermodynamic uncertainty relation applicable to general nonequilibrium dynamics. 

To be concrete, we consider a trajectory of stochastic events across $M$ steps: $[x] := (x_1, ...,x_M)$, whose distribution is given by $P([x])$, and let $[x]^*:=(x_M, ...,x_1)$ be its time-reversed trajectory under the time-reversed protocol, whose distribution is denoted as $P^*([x]^*)$. Each $x$ can be a set of real numbers if the relevant dynamics occurs in a continuous space or a set of discrete values if the space is discrete \footnote{Also, to be concrete, we have specified $[x]$ to be a series of events in discrete time; we note that one can replace $[x]$ by a continuous-time trajectory and the sums by path integrals to show that in continuous time, our results remain unchanged.}. We denote the average with respect to $P([x])$ as $\langle \cdot \rangle$, and that with respect to $P^*([x])$ as $\langle \cdot \rangle_{\rm rev}$. By the Markovian property, the trajectory probability $P([x])$ factorizes into a product of the initial distribution $P_0(x_0)$ and the transition probabilities
\[P([x]) =  P(x_M|x_{M-1}, t_{M-1})...P(x_1|x_0, t_{0})P_0(x_0), \]
where the transition probabilities are explicitly labeled with $t_i$ to stress that the dynamics can be time-dependent.

It has recently been shown \cite{barato2016cost, liu2020thermodynamic, Seifert2019, hasegawa2019fluctuation, shiraishi2016universal, pietzonka2018universal, ito2020stochastic} that the average and variance of any thermodynamic quantity are related to the entropy production through the thermodynamic uncertainty relation (TUR):
\begin{equation}\label{eq: intro tur}
     G[\Delta S]  \geq \frac{\langle J\rangle^2}{\text{Var}[J]},
\end{equation}
where $G[\cdot]$ is a functional of entropy production $\Delta S:= - \log \frac{P^*([x]^*)}{P([x])}$, and $J=J([x])$ is an antisymmetric current, which is odd under time reversal: $J([x])=-J([x]^*)$. A physical interpretation of \eqref{eq: intro tur} is that the relative accuracy of a measurement $\langle J\rangle^2/\text{Var}[J]$ is bounded from above by the entropy production \cite{falasco2020unifying}. A crucial observation here is that a more accurate measurement can be performed only at the cost of higher entropy production.

However, the original TUR only holds in a linear-response regime and is applicable to thermodynamic currents. Various attempts at generalizing the TUR have been made \cite{proesmans2017discrete, garrahan2017simple, horowitz2020thermodynamic, hasegawa2019fluctuation, koyuk2020thermodynamic, liu2020thermodynamic, pal2021thermodynamic, timpanaro2019thermodynamic, barato2018bounds, francica2022fluctuation, potts2019thermodynamic, vroylandt2020isometric}. Notably, Ref.~\cite{hasegawa2019fluctuation} derives a TUR from the fluctuation theorem and the derived TUR is applicable to observables that are not limited to currents; however, it can only be applied to systems with strong constraints on the initial and final states of the system. Reference~\cite{liu2020thermodynamic} generalizes the TUR to an arbitrary nonequilibrium initial state, where the bound only applies to the boundary value of the current. One of the most general forms of existing TUR proposed in Ref.~\cite{dechant2020fluctuation} is applicable to an arbitrary reversible system and an arbitrary observable. However, this relation is still limited in the scope of applicability because it cannot be extended to the situations where the absolute irreversibility is involved \footnote{The right-hand side of the proposal of \cite{dechant2020fluctuation} contains the term $\langle \Delta S\rangle$ explicitly, which may diverge when the system is not fully reversible. The relation in \cite{dechant2020fluctuation} thus becomes a trivial relation: $\infty\geq g(f)$, where $g$ is a function of the relevant observable $f$, whereas our result always remain meaningful. For detailed works about absolute irreversibility, see, for example, Ref. \cite{ness2020absorbing, pal2021thermodynamic, murashita2014nonequilibrium, pietzonka2021classical}.}. %Systems with absolute irreversibility constitute a very important class of systems that appear both within and outside physics. Within physics, free expansion involves absolute irreversibility and appears both in thermodynamics and cosmology.

%The conventional TUR suffers two main limitations. Firstly, the entropy production diverges when either the forward probability $P([x])$ or the backward probability $P^*([x]^*)$ vanishes for some trajectory \cite{proesmans2017discrete, horowitz2020thermodynamic, hasegawa2019fluctuation, koyuk2020thermodynamic}, which makes the conventional TUR inapplicable to situations in which the initial state is far from equilibrium, absorbing states exist, or the underlying process relies on chemical catalysis whose reverse process is forbidden \cite{ness2020absorbing, pal2021thermodynamic, murashita2014nonequilibrium, pietzonka2021classical}. Secondly, the previous TURs often only apply to time-antisymmetric observables \cite{liu2020thermodynamic, pal2021thermodynamic, hasegawa2019fluctuation}, which are often thermodynamic observables in nature. These limitations hinder our understanding of a general nonequilibrium process. Attempts to unify and generalize various TURs have been made. For example, Refs.~\cite{falasco2020unifying} and \cite{koyuk2020thermodynamic} propose unifying TURs by proposing a common form of TURs, but the crucial link to the master fluctuation theorem has remained elusive, and the obtained TURs are thus less general than our result. Two other attempts to derive TURs from the fluctuation theorem can be found in Refs. \cite{hasegawa2019fluctuation} and \cite{timpanaro2019thermodynamic}. However, these TURs can only be applied to restricted settings due to their reliance on Crook's fluctuation theorem (FT) and their goal of studying an antisymmetric observable.

In this Letter, we overcome the limitations of the previous generalizations of the TUR and derive a universal TUR that is applicable to an arbitrary system with or without absolute irreversibility and to an arbitrary observable. Also, our result only involves the quantities that appeared in the original TUR. We prove the following universal inequality for an arbitrary observable $f=f([x])$ \footnote{See Appendix A for the proof.}:
\begin{equation}\label{eq: main result intro}
    \frac{\langle e^{2R} \rangle}{\gamma^2}  \geq 1 +  \frac{(\langle f \rangle_{\rm Q:P>0} - \langle f\rangle)^2}{\text{Var}[f] },
\end{equation}
where $R:=\log Q/P$ is the log-probability ratio between the probability of the primary dynamics $P$ and a reference dynamics $Q$, and $\langle f \rangle_{Q:P > 0}:= \mathbb{E}_Q[f|P([x])>0]$ is the conditional expectation value of $f$ under $Q$, conditioned on $P([x])>0$. The freedom of choice of the reference dynamics is a generic feature of fluctuation theorems \cite{murashita2014nonequilibrium, esposito2010three}. The choice with physical relevance is $Q=P^*([x]^*)$, which makes the left-hand side of the inequality dependent on the entropy production: $e^{2R}=e^{-2\Delta S}$ \cite{murashita2014nonequilibrium, seifert2012stochastic}. The term $\gamma:=\langle e^{-\Delta S}\rangle$ on the left-hand side represents the degree of reversibility in the system, which is equal to $1$ (0) when it is fully reversible (irreversible) \cite{murashita2014nonequilibrium}. 

%We first discuss the difference between \eqref{eq: main result intro} and the previous versions of TUR. The most standard form of TUR discussed in Ref.~\cite{horowitz2020thermodynamic} is only applicable to systems in nonequilibrium steady states. The form of TUR in Ref.~\cite{hasegawa2019fluctuation} is closer in form to ours in that it is derived from the Crook fluctuation theorem; however, it requires the protocol to be symmetric with respect to time reversal and cannot be applied to arbitrary initial and final states. The TUR in Ref.~\cite{proesmans2017discrete} requires the protocol to be independent of time. A form of TUR that applies to arbitrary initial and final states is proposed in Ref.~\cite{liu2020thermodynamic}; however, the result in this work can only be applied to systems with a time-independent protocol, antisymmetric currents, and the detailed balance condition. Reference~\cite{pal2021thermodynamic} presents the only other type of TUR that applies to a system with absolute irreversibility. However, it can only be applied to antisymmetric observables. Reference~\cite{koyuk2020thermodynamic} is the only other TUR that also applies to arbitrary initial states and arbitrary time-dependent protocols, but it requires the transition rates to satisfy the local detailed balance condition and thus does not apply to systems outside physics.

\textit{General Properties of the Universal TUR}. If the reference dynamics is the time-reversed one of the forward dynamics, i.e., $Q=P^*([x]^*)$, then our TUR \eqref{eq: main result intro} leads to the following nontrivial relation between the current $f$, its variance $\text{Var}[f]$, the degree of absolute irreversibility $\gamma$, and the total entropy production $\Delta S$:
\begin{equation}\label{eq: entropy tur}
    \text{Var}[f] \geq \frac{(\langle f \rangle_{\rm P^*([x]^*):P>0} - \langle f\rangle)^2}{\langle e^{-2\Delta S} \rangle/\gamma^2  -1}.
\end{equation}
This TUR is universal in that it applies to arbitrary initial and final states, allows for arbitrary time-dependent protocols, can be applied to arbitrary observables, makes no assumption about the transition probabilities, and is applicable to both continuous-time and discrete-time dynamics. The generality of the derived TUR is a consequence of the generality of the master fluctuation theorem, which finds its mathematical foundations in the general change-of-measure theorem in the measure theory \cite{murashita2014nonequilibrium}. A key feature of the this bound is that it involves an exponentiated entropy production, which may be dominated by rare trajectories \cite{jarzynski2006rare}. However, we note that this is not a weakness of the proposed theory, but a reflection of the underlying physics that rare trajectories can have significant influence on the expected strength and fluctuation of physical quantities. This argument is further substantiated by the fact that no matter how strongly the term $\langle e^{-2\Delta S} \rangle$ is dominated by the rare trajectories, there is always some observable that makes the bound satisfied.   %Our result, therefore, indeed presents a significant advancement in understanding the general nonequilibrium phenomena. 

The choice of $Q$ suitable for the system and observable is crucial. For example, the entropy term $e^{2R}$ is minimized when the reference dynamics $Q$ is as close to the original one as possible; the current term $\langle f \rangle_{\rm Q:P>0} - \langle f\rangle$ is maximized when $Q$ is chosen to make $\langle f \rangle_{\rm Q:P>0}$ have a sign opposite to that of $\langle  f \rangle$. We will see later that appropriate choices of $Q$ lead to meaningful results for physics and theoretical finance.
Also, the result in \eqref{eq: main result intro} achieves two types of optimality. Firstly, for every system, there exists an observable $f$ such that the bound is saturated. Secondly, for every observable $f$, there exists $P$ and $Q$ such that the bound is saturated \footnote{See Appendix B for the proof.}. These optimalities imply that having a different form of TUR leads to either a looser bound for some systems or making the bound inapplicable to some observables. For example, the standard TUR has a linear entropic term \cite{horowitz2020thermodynamic} in place of $e^{-2\Delta S}/\gamma^2-1$ in our bound with $\Delta S$. However, it is not hard to see that the standard TUR bound is trivial for any irreversible process: when the support of $P^*$ is a proper subset of that of $P$, $ \Delta S  = \infty$, so the left-hand side of the standard TUR \eqref{eq: intro tur} diverges, and we obtain the trivial result $\text{Var}[f] \geq 0$. Some bounds are more similar to ours and involve an exponential term, but with the plus sign in the exponent \cite{proesmans2017discrete, hasegawa2019fluctuation}: $e^{\Delta S}$; this type of formula also cannot appear in the most general TUR because $e^{\Delta S}$ also diverges when $\gamma \neq 1$. Among the three choices of the entropic term ($\Delta S$, $e^{-\Delta S}$, $e^{\Delta S}$), only the form $e^{-\Delta S}$ can remain finite. %Lastly, we note that the fact our result in \eqref{eq: main result intro} is applicable to all observables offers a potential answer to the fact that the FTTURs are often much looser than the conventional TURs in bounding a time-antisymmetric current \footnote{Inequality~\eqref{eq: main result intro} belongs to the family of FTTURs (the TURs derived from FTs), which is sometimes criticized for being looser than the standard TUR in bounding antisymmetric currents \cite{liu2020thermodynamic, thibado2020fluctuation}. However, the reason for its looseness is not well understood. Our result offers two explanations. Firstly, the FTTURs can be applied to arbitrary quantities, not limited to antisymmetric current, which explains why FTTURs are looser when we restrict ourselves to antisymmetric currents. Secondly, we show in Appendix A that our derived FTTUR is optimal for the class of observables that share the same sign (See Theorem 1), which explicitly excludes antisymmetric currents that take on the opposite sign when the trajectory is reversed.}.

Inequality~\eqref{eq: main result intro} takes a simpler form for antisymmetric currents. Let the quantity $f$ be an antisymmetric current, the protocol be time-independent and the initial and final distributions are stationary, our TUR reduces to: $\langle e^{ \Delta S} \rangle  - 1 \geq \frac{ 4\langle f\rangle^2}{\text{Var}[f] }$ \footnote{See Appendix D}. Expanding this relation to first order in $\Delta S$, it recovers the original TUR in Ref.~\cite{barato2016cost}, which holds in the linear-response regime.
%We stress that this bound is not tighter than the standard TUR when restricted to the antisymmetric currents because the left-hand side can still upper bound many observables that are not antisymmetric \footnote{The derivation and a few other intermediate cases are given in Appendix C.}. 
Also, as in Ref.~\cite{timpanaro2019thermodynamic}, we can also generalize the main result in \eqref{eq: main result intro} to a vector-valued observable $\mathbf{f}$ as detailed in Appendix C, which is the most general TUR we derive. Lastly, we note that the proposed relation also takes a meaningful form in the equilibrium limit where $\Delta S$ approaches zero. We study this in Appendix F.

\textit{Application I -- Interplay between the degree of non-stationarity and entropy production}. {When the transition probabilities are time-independent and $\gamma=1$, our result offers a crucial insight into the achievable measurement accuracy of a thermodynamic current $f$ that is antisymmetric against time reversal. Let $P_0$ and $P_M$ denote the initial and final distributions of the process $P([x])$. We choose $Q$ to be the distribution resultant from the time-reversed dynamics, with $P_0$ as the initial distribution: $Q=P^*([x]^*)P_0(x_M)/P_M(x_M)$. Then, ~\eqref{eq: main result intro} becomes
\begin{equation}\label{eq: our antisymmetric TUR}
    \left\langle e^{-2\Delta S - 2D(P_M||P_0)} \right\rangle  -1 \geq \frac{4\langle f\rangle^2}{\V[f]},
\end{equation}
where $D(P_M||P_0):=\log \frac{P_M(x_M)}{P_0(x_M)}$ \footnote{We clarify that $D$ is not equal to the system-wise entropy change. Following Ref~\cite{crooks1999entropy}, the trajectory-wise system entropy change is $D(P_M(X_M)|| P_0(x_0))$, whereas the $D$ term in our bound is $D(P_M(x_M)||P_0(x_M))$. The two are different in general.}, $\langle D\rangle$ is the Kullback-Leibler divergence between the initial and final distributions. We see that $D$ measures the trajectory-wise distance between the initial distribution and the final (possibly non-stationary) distribution. The right-hand side of \eqref{eq: our antisymmetric TUR} is the relative accuracy. Noting that $\langle D\rangle$ is zero when the system is stationary, it characterizes the ``degree of non-stationarity." Thus the interplay between the degree of non-stationarity and entropy production plays a key role in providing the accuracy of a thermodynamic current in the most general Markovian relaxation dynamics, whose initial and final distributions can be out-of-equilibrium. Specifically, we find that the maximum achievable accuracy decreases with an increasing deviation between the initial and final distributions. Physically, this strong dependence on the initial condition can be understood as follows: when measuring a local parameter close to a site $z$, it is the most efficient for us to initialize the state of the system in the neighborhood of $z$. If we choose the initial state according to the Boltzmann distribution, many states away from $z$ are involved, resulting in a reduced measurement efficiency.

The conventional TUR dictates that the bound on the measurement accuracy of antisymmetric currents increases as we increase the entropy production \cite{seifert2012stochastic, falasco2020unifying}. On the contrary, our result implies that the measurement accuracy can decrease with an increasing entropy production when the system is out-of-equilibrium and when $D$ dominates the entropy production. Lastly, when the system is close to stationarity, $D$ becomes negligible, and one can show that the bound approximately reduces to $\langle e^{ \Delta S} \rangle  - 1 \geq \frac{ 4\langle f\rangle^2}{\text{Var}[f] }$, which, in agreement with the standard TUR, suggests that the limit of measurement accuracy should increase with the entropy production.

\begin{figure}
    \centering
    \includegraphics[width=0.95\linewidth]{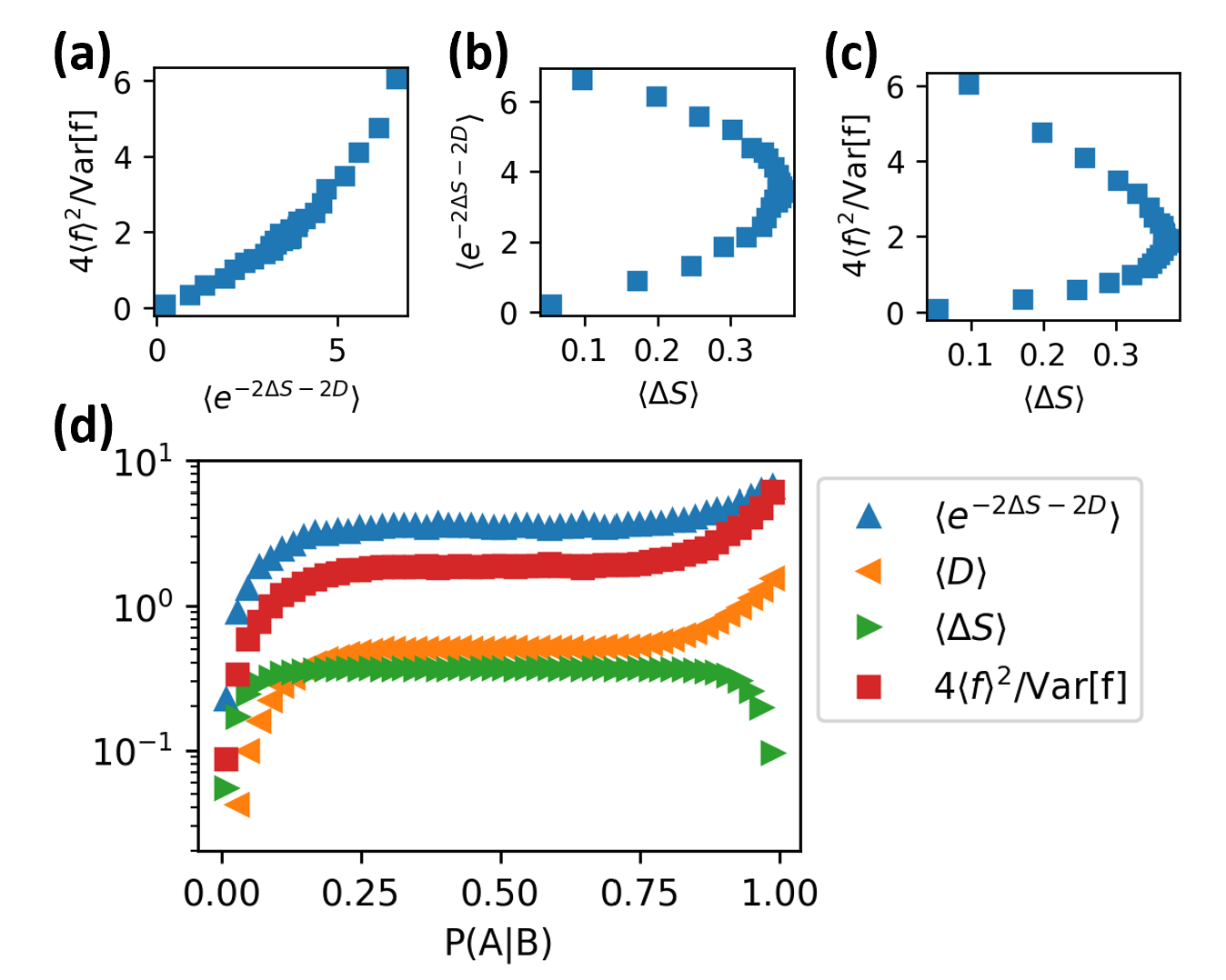}
    \vspace{-1.2em}
    \caption{Numerical simulation of a two-state system, where an antisymmetric current is measured. (a) Accuracy vs. the proposed bound in \eqref{eq: our antisymmetric TUR}. The proposed bound agrees with the measurement accuracy in both trend and magnitude. (b) Proposed bound vs. $\langle \Delta S\rangle$. $\langle \Delta S\rangle$ both increases and decreases with the bound. (c) Accuracy vs. $\langle \Delta S\rangle$. $\langle \Delta S\rangle$ both increases and decreases with accuracy. (d) All relevant quantities together. For a wide range of transition probabilities, we see that $D$ is comparable to or dominates $\Delta S$, showing that it does play an important role in the bound.}
    \vspace{-1em}
    \label{fig:accuracy-entropy curve}
\end{figure}
As an example, we numerically simulate an out-of-equilibrium two-state system. The two states are labeled as $A$ and $B$. We let the initial state be $P_A(0) = 0.9$ and $P_B(0) = 0.1$. The transition probability is set to be symmetric: $P(A|B) = P(B|A)$. We study the bound at different $P(A|B)$, varying from zero to one. The observable $f$ we consider is the net number of transitions from state $A$ to state $B$: $f:= \delta_{x_{M}, B} - \delta_{x_0, A} $, which is by definition an antisymmetric observable. Figure~\ref{fig:accuracy-entropy curve} shows the relevant quantities of this simulation. We see that the proposed relation~\eqref{eq: our antisymmetric TUR} places a bound of the measurement accuracy of the far-from-equilibrium system across all transition probabilities, whereas the standard TUR is almost everywhere inapplicable. A detailed comparison is given in Appendix E, where we show that the bound in inequality~\eqref{eq: our antisymmetric TUR} can be considerably tighter than other TURs.

\textit{Application II -- A Fundamental Bound in Theoretical Finance}. This example shows that it is, in fact, important to freely choose $Q$ if we want to make the TUR relevant to general nonequilibrium dynamics in fields other than physics. The financial market is a major non-physical nonequilibrium system that impacts our daily life \cite{sornette2014physics, stanley2008statistical}. Consider the price trajectory of a product (e.g., a stock) that changes from time $0$ to time $\tau$, denoted as $x_0,...,x_\tau$, where each unit of time corresponds to a day. The price return is defined as $r_t = \frac{x_{t+1} - x_t}{x_t}$. Here, we assume that the price dynamics follows a discrete-time Markovian dynamics in continuous space. While this assumption may not hold in general, it holds for the standard minimal models in finance such as the Black-Scholes model \cite{black2019pricing} and the Heston model \cite{heston1993closed}. A key quantity is the volatility of the price return $\sigma := \sqrt{\text{Var}[r]}$.

With $Q=P^*([x]^*)$, our TUR in \eqref{eq: main result intro} gives
\begin{equation}\label{eq: entropy bound for finance}
    \sigma \geq \sqrt{\frac{(\langle  r\rangle - \langle  r^*\rangle_{\rm rev})^2}{\langle e^{-2\Delta S} \rangle - 1}},
\end{equation}
which explicitly shows that the thermodynamic entropy production gives a bound on price volatility, which can be useful for problems such as option pricing. This may open a venue for studying quantitative finance in terms of stochastic thermodynamics. 
Moreover, since the price return is not a time-antisymmetric observable, the standard TUR cannot apply.

We now show that a fundamental inequality in theoretical finance can be derived as a special case of the general TUR in \eqref{eq: main result intro}. By investing a fraction of one's capital, $p_t$, in the product at different times, one can make profit, which is measured with the wealth return rate $\mathcal{R}_t$: $\mathcal{R}_t(p) := \frac{x_{t+1} - x_t}{x_t}p + r_f (1-p)$, where $r_f$ is the risk-free interest rate. In theoretical finance (capital asset pricing model), a fundamental quantity is the Sharpe ratio, defined as
\begin{equation}
    \chi(p) := \frac{\langle \mathcal{R} \rangle -r_f }{\sqrt{\text{Var}[\mathcal{R}]}},
\end{equation}
where $\mathcal{R} := \prod_{i=0}^{\tau-1} (1 + \mathcal{R}_{i}) -1$. Again, the observable $\mathcal{R}$ is not antisymmetric, and therefore the standard TURs do not apply. The Sharpe ratio is widely accepted as a fundamental quantity in theoretical finance \cite{sharpe1966mutual, sharpe1964capital} and used in practice as a metric of successful investment.
Theoretically, it is known that optimal portfolios should all have the same maximized Sharpe ratio \cite{sharpe1966mutual}, and it is an important problem to find an upper bound on the Sharpe ratio.  

For this problem, the relevant reference dynamics is no longer $P^*([x]^*)$ because $\langle r^* \rangle_{\rm rev} \neq r_f$ in general. We need to choose $Q$ to be the dynamics such that the wealth grows at the risk-free rate on average. This dynamics can be achieved if the stock price grows deterministically as $x_t = (1+r_f)x_{t-1}$, or if we create and transact a financial derivative according to the Black-Scholes formula. With $f=\mathcal{R}$, our TUR yields $\V[\mathcal{R}] \geq \frac{(\langle \mathcal{R} \rangle - r_f)^2 }{\langle e^{2R} \rangle -1}$, or, equivalently,
\begin{equation}\label{eq: sharpe ratio bound}
    \sqrt{\langle e^{2R} \rangle -1} \geq \chi(p),
\end{equation}
for any trading strategy. The existence of such a $Q$ is guaranteed by the fundamental theorem of finance (no-arbitrage theorem), and inequality~\eqref{eq: sharpe ratio bound} applies to any price dynamics that obeys the fundamental theorem of finance. In fact, this bound has the same form as the celebrated Hansen-Jagannathan (HJ) bound in theoretical finance \cite{hansen1991implications,  cochrane2000beyond, bjork2006towards}. The HJ bound is a fundamental relation in theoretical finance because it applies to all models of the market, and there have been many applications of it besides upper bounding the Sharpe ratio \cite{kan2006new, snow1991diagnosing, stutzer1995bayesian, gagliardini2020comparing}. While the original HJ bound can only be applied to the case in which $Q$ is a martingale distribution, our result applies to an arbitrary distribution $Q$ such that $\langle \mathcal{R}\rangle_Q = r_f$. %Our bound potentially gives a tighter bound on the achievable profitability than the original HJ bound because the space of all possible distributions is much larger than the space of martingale distributions. %A numerical verification of our version of the bound is provided in Appendix~\ref{app sec: finance numerical}. 
The bound~\eqref{eq: entropy bound for finance} is a new relation we discovered. One important utility of the proposed relation for finance is that it allows one to check the validity and correctness of existing theories of finance (this is also what the related Hansen–Jagannathan bound can be useful for). For example, existing stock price data allows one to estimate the return $r$ and its volatility $\sigma$, and the minimal models allow one to calculate the time-reversed return $r^*$, and the entropic term $e^{-2\Delta S}$, and this can be plugged into the proposed TUR relation, the violation of which can then be used to reject the economic theory under consideration. Our result thus offers a novel method to test the validity of economic theories with physics-relevant quantities (such as the entropy production rate), which presents yet another remarkable usage of physics principles in other fields.

This application shows that the HJ bound in finance is comparable to the thermodynamic uncertainty relations in physics, and the crucial difference between the two bounds arises from the choice of the reference probability $Q$. The choice of fundamental importance in physics is the time-reversed dynamics $P^*([x]^*)$, while the fundamental choice in finance is the martingale measure, under which one obtains the risk-free return. Therefore, our derived TUR unifies the fundamental bounds in physics and finance.

\textit{Conclusion and discussion}. We have derived a universal form of TUR for an arbitrary Markovian dynamics in discrete space-time, which includes the continuous space-time dynamics as a special limit. Our bound is shown to achieve two kinds of optimalities, but it is unlikely to be the optimal bound if we restrict the problem to systems of special types or observables with specific symmetries. Investigating how such constraints on observables may help improve the bound is an important future work. One crucial quantity identified in this work is the degree of non-stationarity $D$, and it should be important to understand it better in the future. Our result also links the fundamentals in theoretical finance and physics, and further exploring this connection may lead to exciting cross-fertilization of both fields.

\begin{acknowledgments}
This work was supported by a KAKENHI Grant No. JP18H01145 from the Japan Society for the Promotion of Science. Ziyin thanks Tonghua Yu and Junxia Wang for many thoughtful promenades they shared during the writing of this paper.
\end{acknowledgments}

%\bibliography{ref}


% Produces the bibliography via BibTeX.

\end{document}

\title{Appendix for ``Universal Thermodynamic Uncertainty Relation in Non-Equilibrium Dynamics"}
\author{Liu Ziyin$^1$}
\author{Masahito Ueda$^{1,2,3}$}%
\affiliation{%
$^1$Department of Physics, The University of Tokyo, 7-3-1 Hongo, Bunkyo-ku, Tokyo 113-0033, Japan\\
$^2$RIKEN Center for Emergent Matter Science (CEMS), Wako, Saitama 351-0198, Japan \\
$^3$Institute for Physics of Intelligence, The University of Tokyo, 7-3-1 Hongo, Bunkyo-ku, Tokyo 113-0033, Japan
}%

\date{\today}
\maketitle

\onecolumngrid
\appendix

%\textit{Derivation of Inequality~\eqref{eq: main result intro}}. 

We first note that for notational consistency, we treat the expectations as a sum over discrete trajectories:
\begin{equation}
    \langle f\rangle = \sum_{[x]} P([x]) f([x]).
\end{equation}
One can simply replace the sum by path integrals to obtain the same results for continuous paths.

\section{Derivation of the Main result}
Let $R:= \log P^*([x]^*)/P([x])$. For an arbitrary observable $f = f([x])$ and a reference dynamics $Q$, the master fluctuation theorem with absolute irreversibility \cite{murashita2014nonequilibrium} implies
\begin{align}\label{eq: non equilibirum identity}
&\gamma \langle f \rangle_{Q:P([x])> 0} = \left\langle f\exp( R)\right\rangle_{P},
\end{align}
where $\gamma := \sum_{[x]: P([x])>0} Q([x]) \leq 1 $, and 
\begin{equation}
    \langle f \rangle_{Q:P([x])> 0} :=  \frac{1}{\sum_{[x]: P([x])>0} Q([x])}\sum_{[x]: P([x])>0} Q([x]) f([x]).
\end{equation}
Note that when setting $f=1$, one obtains the integral fluctuation theorem with absolute irreversibility: $\langle e^{R} \rangle =\gamma$.

The Cauchy-Schwarz inequality gives $\text{Var}[f] \text{Var}[\exp(R)] \geq \text{Cov}( f,\exp(R))^2,$ where
\[    \text{Var}[\exp(R)]= \langle \exp(2R) \rangle  - \langle \exp(R) \rangle^2 = \langle \exp(2R) \rangle  - \gamma^2,
\]
which follows from the integral fluctuation theorem: $\langle e^R \rangle = \gamma$. The master FT implies that $\text{Cov}(f,\exp(R))= \gamma(\langle f \rangle_{\rm Q:P>0} - \langle f\rangle)$. We thus have the main result of this work:
\begin{equation}\label{eq: generalized TUR form 2}
    \text{Var}[f] \geq \frac{(\langle f \rangle_{\rm Q:P>0} - \langle f\rangle)^2}{\langle e^{2R} \rangle/\gamma^2  - 1},
\end{equation}
which is equivalent to (2) in the main text. The right-hand side of (2) includes the term $\frac{(\langle f \rangle_{\rm Q:P>0} - \langle f\rangle)^2}{\text{Var}[f] }$, which can be interpreted as a general form of the measurement accuracy of $f$ relative to a reference point $\langle f \rangle_{\rm Q:P>0}$ \footnote{Recent advances in stochastic thermodynamics suggest that the term $\langle f\rangle$ cannot appear in the TUR when considering the dynamics beyond a near-equilibrium steady state (NESS), and some additional term, often in the form of generalized biases or boundary currents, also needs to appear in the numerator. Therefore, the additional term in our bound agrees with the recent results that generalize the TUR beyond the NESS setting \cite{pal2021thermodynamic, liu2020thermodynamic, koyuk2020thermodynamic}}.

\section{Optimality}\label{app sec: optimality proof}
As mentioned in the main text, there are two types of optimalities that the proposed relation achieves:
\begin{enumerate}
    \item for every system (any $P$ and $Q$), there exists an observable $f$ such that the bound is saturated;
    \item for every observable $f$, there exists $P$ and $Q$ such that the bound is saturated.
\end{enumerate}
In essence, both optimalities are proved by the fact that the Cauchy-Schwarz inequality is tight when (and only when) the two relevant random variables are proportional to each other.

To prove the first optimality property, simply let $f= e^R$, and one can then show that equality holds. To focus on the physics relevance, we let $R = -\Delta S$. The right-hand side of (2) becomes $\langle e^{2R} \rangle/\gamma^2  - 1$ and so equality holds.

To prove the second optimality property, we first assume without loss of generality that the observable under consideration $f([x])$ is non-negative for some $[x]$ (otherwise, just consider the non-positive part in a similar manner). We first choose an arbitrary distribution $P$ such that the following properties hold:
\begin{enumerate}
    \item $\langle f \rangle_P$ is finite;
    \item $P([x])> 0$ if and only if $f([x]) \geq 0$.
\end{enumerate}
Now, let $Q([x]) = \frac{1}{\sum f([x])P([x])} f([x])P([x])$, which exists because $\langle f \rangle_P$ is finite. One can then employ this choice of $P$ and $Q$ to show that the bound is saturated. 

Note that $Q$ and $P$ have the same support, and so $\gamma =1$, the proposed bound thus reads
\begin{equation}\label{eq: optimality proof bound}
    {\rm{Var}}[f] \geq \frac{(\langle f\rangle - \langle f \rangle_Q)^2}{\langle e^{2R}\rangle -1 }.
\end{equation}
Each quantity can be directly found
\begin{equation}
    \langle f \rangle_Q = \sum Q f = \frac{1}{\sum f P} \sum f^2 P = \frac{\langle f^2 \rangle}{\langle f\rangle}.
\end{equation}
\begin{equation}
    \langle e^{2R}\rangle = \sum P \frac{Q^2}{P^2} = \frac{1}{(\sum fP)^2} \sum P \frac{(f P )^2}{P^2} = \frac{\langle f^2 \rangle}{\langle f\rangle^2}.
\end{equation}
Substitute into \eqref{eq: optimality proof bound}, one sees that the equality is achieved.

\section{Matrix TUR}\label{proof: matrix tur}

This matrix form bound can be tighter than inequality (2) when one wants to simultaneously bound two correlated variables.
\begin{theorem}
    %Let $\langle \mathbf{f}\rangle:=\langle \mathbf{f}\rangle$, $\langle \mathbf{f}\rangle_Q:=\langle \mathbf{f}\rangle_{Q:P>0}$ and $C:= \text{cov}_P(\mathbf{f}, \mathbf{f})$. Then
    Let $\mathbf{f}$ be an arbitrary vector observable. Then, 
    \begin{equation}
        \text{cov}_P(\mathbf{f}, \mathbf{f}) \geq  \frac{(\langle \mathbf{f}\rangle_Q - \langle \mathbf{f}\rangle) (\langle \mathbf{f}\rangle_Q - \langle \mathbf{f}\rangle)^{\rm T}}{\langle e^{2R} \rangle/\gamma^2  - 1}.
    \end{equation}
\end{theorem}

\textit{Proof}. We start from the general TUR in (10) for a general scalar observable $f'$:
\begin{equation}\label{eq: app proof 1}
    \text{Var}[f'] \geq \frac{(\langle f' \rangle_{\rm Q:P>0} - \langle f'\rangle)^2}{\langle e^{2R} \rangle/\gamma^2  - 1}.
\end{equation}
Let $f'=f_\mathbf{u}(\mathbf{f}):= \mathbf{u}^{\rm T} \mathbf{f}$ for some constant $\mathbf{u}$. Plug into the left-hand side of \eqref{eq: app proof 1}, we obtain
\begin{align}
    \text{Var}[f'] &= \mathbf{u}^{\rm T} C \mathbf{u}.
\end{align}
The right-hand side of \eqref{eq: app proof 1} reads:
\begin{equation}
    \frac{(\langle f' \rangle_{\rm Q:P>0} - \langle f'\rangle)^2}{\langle e^{2R} \rangle/\gamma^2  - 1} = \mathbf{u}^{\rm T} \frac{(\langle \mathbf{f}\rangle_Q - \langle \mathbf{f}\rangle) (\langle \mathbf{f}\rangle_Q - \langle \mathbf{f}\rangle)^{\rm T}}{\langle e^{2R} \rangle/\gamma^2  - 1} \mathbf{u}.
\end{equation}
\eqref{eq: app proof 1} can thus be written as
\begin{equation}
    \mathbf{u}^{\rm T} C \mathbf{u} - \mathbf{u}^{\rm T} \frac{(\langle \mathbf{f}\rangle_Q - \langle \mathbf{f}\rangle) (\langle \mathbf{f}\rangle_Q - \langle \mathbf{f}\rangle)^{\rm T}}{\langle e^{2R} \rangle/\gamma^2  - 1} \mathbf{u} \geq 0
\end{equation}
for an arbitrary $\mathbf{u}$. This inequality is equivalent to the statement that
\begin{equation}
    C -  \frac{(\langle \mathbf{f}\rangle_Q - \langle \mathbf{f}\rangle) (\langle \mathbf{f}\rangle_Q - \langle \mathbf{f}\rangle)^{\rm T}}{\langle e^{2R} \rangle/\gamma^2  - 1} 
\end{equation}
is positive semi-definite. This completes the proof. $\square$

\section{Other Special Cases of TUR}\label{app sec: intermediate turs}

Let the quantity $f$ be an antisymmetric current as in the standard TURs: $f=\sum_{i=0}^{M-1} g(x_i, x_{i+1})$
for antisymmetric functions $g(x_i, x_{i+1}) = -g(x_{i+1}, x_{i})$. In this case, it is easy to check that $f^*= - f$. The TUR takes the form \footnote{It might be helpful to note that $\langle f\rangle_{P^*([x]^*)} =\langle f^*\rangle_{P^*([x])} = \langle f^*\rangle_{\rm rev} $.}:
\begin{equation}
    \text{Var}[f] \geq \frac{(\langle f \rangle_{{\rm rev, P>0}} +  \langle f\rangle)^2}{\langle e^{-2 \Delta S} \rangle/ \gamma^2 -1}.
\end{equation}
From now on, we focus on the case of $\gamma=1$. The TUR then becomes:
\begin{equation}
    \text{Var}[f] \geq \frac{(\langle f \rangle_{{\rm rev}} + \langle f\rangle)^2}{\langle e^{-2 \Delta S} \rangle  - 1}.
\end{equation}
When the protocol is time-independent, the reversal dynamics is the same as the forward dynamics and so $\langle f\rangle_{\rm rev}$ is nothing but the expectation of $f$ running from time $\tau$ to time $2\tau$. Therefore, we obtain a simpler form:
\begin{equation}
    \text{Var}[f] \geq \frac{ (\langle f\rangle_{\rm rev}+ \langle f\rangle)^2}{\langle e^{-2 \Delta S} \rangle  - 1} = \frac{ (\langle f(2\tau)\rangle)^2}{\langle e^{-2 \Delta S} \rangle  - 1}.
\end{equation}
This shows that the fluctuation of an antisymmetric current after time period $\tau$ is bounded from below by its expected value at time $2\tau$.

However, \textit{prima facie}, this inequality seems to contradict the standard TURs: the denominator at the right-hand side ($e^{-2 \Delta S}$) seems to decrease as the entropy production increases. That is to say, the quantity $f$ becomes easier to measure as the entropy production increases: this is the opposite tendency to that of the standard TURs, where the measurement of $f$ becomes harder as the entropy production decreases. However, a closer examination suggests that the proposed relation is consistent with the standard TURs. In fact, the proposed relation can reduce to a form similar to the standard TUR. To show this, let $f_1 = e^{- \Delta S}$. Then, 
\begin{equation}
    \langle e^{-2 \Delta S} \rangle = \langle f_1 e^{- \Delta S} \rangle  = \langle f_1^* \rangle_{{\rm rev}}.
\end{equation}
By the definition of $f_1=e^{- \Delta S}$, we have $f^*_1 = e^{ \Delta S}$, i.e.,
\begin{equation}
    \langle e^{-2 \Delta S} \rangle = \langle e^{\Delta S}\rangle_{{\rm rev}}.
\end{equation}
Note that this relation holds for any system such that the main theorem is applicable. This leads to the following general inequality:
\begin{equation}
    \text{Var}[f] \geq \frac{\langle f(2\tau)\rangle^2}{\langle e^{ \Delta S} \rangle_{\rm rev}  - 1}.
\end{equation}
When the protocol is time-independent and when the initial state is steady, we have that $\langle \cdot \rangle_{{\rm rev}} = \langle\cdot \rangle$ and $f(2\tau) = 2f(\tau)$, and so, for steady-state currents, we have
\begin{equation}
    \text{Var}[f] \geq \frac{ 4\langle f\rangle^2}{\langle e^{ \Delta S} \rangle  - 1}.
\end{equation}
%, with a better constant of $4$. In fact, it has been shown in \cite{falasco2020unifying} that this factor of $4$ is indeed the optimal constant for the steady-state setting, this further corroborates our motivation that establishing the TURs from more fundamental relations is beneficial. 

\subsubsection{An alternative Derivation}
A key step in the above derivation is that when the system is stationary, 
\begin{equation}
    \langle e^{-2\Delta S}\rangle = \langle e^{\Delta S}\rangle.
\end{equation}
This relation can be derived in a more straightforward manner:
\begin{align}
    \langle e^{-2\Delta S}\rangle &= \sum_{[x]}P([x])\left(\frac{P^*([x]^*)}{P([x])}\right)^2\\
    &= \sum_{[x]}P^*([x]^*)\frac{P^*([x]^*)}{P([x])}\\
    &= \sum_{[x]^*}P^*([x]^*)\frac{P^*([x]^*)}{P([x])}  \quad\quad (= \left\langle e^{\Delta S} \right\rangle_{\rm rev})\\
    &= \sum_{[x]^*}P([x]^*)\frac{P([x]^*)}{P([x])}  = \left\langle e^{\Delta S} \right\rangle,
\end{align}
where we have used the fact that the reversed trajectory probability is equal to the forward trajectory probability if the system is stationary: $P^*([x]^*) = P([x^*])$.

\begin{figure}[b!]
    \centering
    \includegraphics[width=0.5\linewidth]{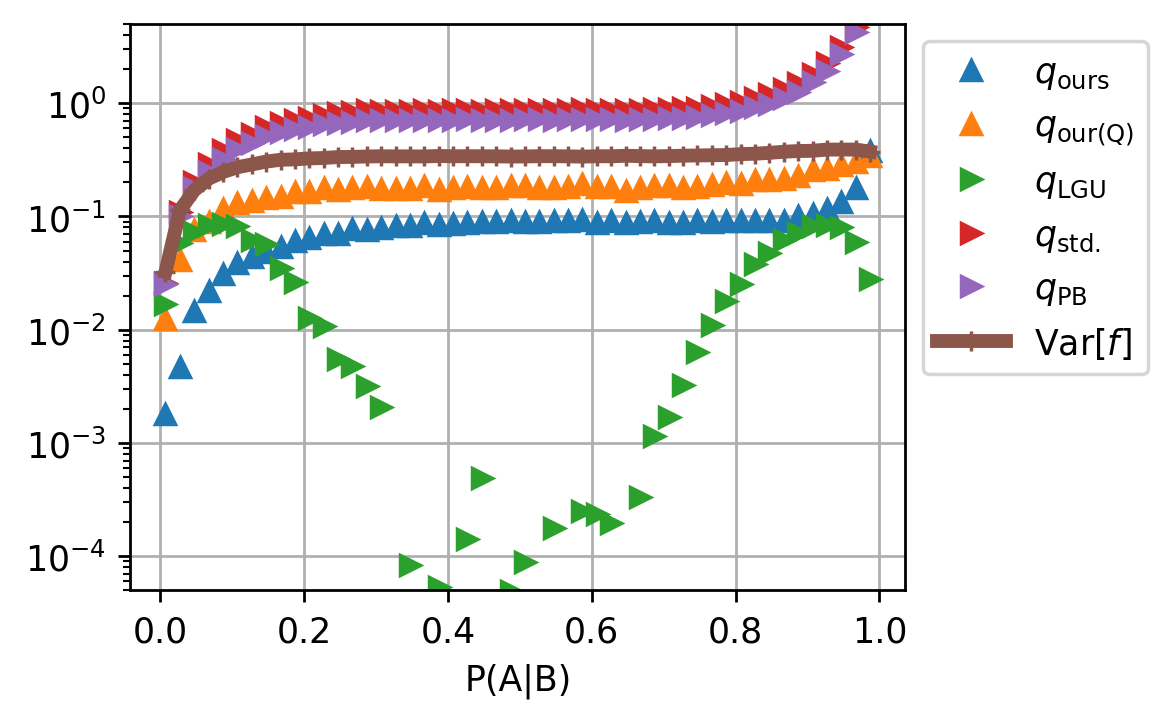}
    \caption{Comparison with the conventional TURs. We see that the standard forms of TUR do not hold due to a non-stationary initial state. The result in \cite{liu2020thermodynamic} holds but does not agree in trend with the actual fluctuation. In contrast, our result can be much tighter and agree in trend, which is a signature that the proposed theory captures the correct essential physics of the dynamics.}
    \label{fig: comparison}
\end{figure}
\section{Numerical Comparison with conventional TURs}\label{app sec: comparison}
In this section, we test the proposed TUR for the cases where conventional bounds apply. A related result is proposed by Ref.~\cite{liu2020thermodynamic}, which also applies to an arbitrary (reversible) initial state for discrete-time dynamics but only for time-independent protocols. Let us first describe again the problem setting.

We consider a $2$-state system with labels $A$ and $B$ with the same energy. We let the initial state be $P_A(0) = 0.9$ and $P_B(0) = 0.1$. The transition probability is set to be symmetric: $P(A|B) = P(B|A)$. We compare the two bounds for varying $P(A|B)$ from zero to one. Note that this system satisfies the detailed balance condition, and the result in \cite{liu2020thermodynamic} indeed applies. All the existing TURs can be seen as a lower bound of the fluctuation of an observable $f$:
\begin{equation}
    \text{Var}[f] \geq q,
\end{equation}
where the term $q$ is different for different TURs, and a TUR can be said to be ``better" if $q$ is closer to $\text{Var}[f]$. We thus make the comparison between the $q$ term of each TUR and $\text{Var}[f]$. 

In this example, the observable $f$ we consider is the net number of transitions from state $A$ to state $B$: $f:= \delta_{x_{\tau}, B} - \delta_{x_0, A} $, which is by definition an antisymmetric observable. To apply the proposed relation (2), we need to specify the reference dynamics. We make two choices: (1) the standard choice $Q=P^*$, and (2) the choice that leads to (4), i.e., $Q=P^*([x]^*)\frac{P_0(x_M)}{P_M(x_M)}$. The difference between these two choices can highlight the advantage of freely choosing $Q$. The reason why this choice is better than the original is that the current term $\langle f\rangle_{\rm rev}$ can be very small for the relaxation process under consideration. Choosing $Q=P^*([x]^*)\frac{P_0(x_M)}{P_M(x_M)}$, on the other hand, makes $\langle f\rangle_{Q}$ comparable to the magnitude of $\langle f\rangle$ and is likely to make the bound much tighter. We show this numerically.

Figure~\ref{fig: comparison} plots the $q$ term from the standard TUR ($q_{\rm std.}$) \cite{horowitz2020thermodynamic} and the discrete-time TUR ($q_{\rm PB}$) \cite{proesmans2017discrete}. We see that both the standard TUR and the discrete-time TUR are not applicable, because they are insufficient to characterize a non-stationary initial state. In contrast, we see that both the result of this work ($q_{\rm ours}$ and $q_{\rm ours(Q)}$) and that of Ref.\cite{liu2020thermodynamic} ($q_{\rm LGU}$) hold as expected. Here, $q_{\rm ours}$ denotes the standard choice and $q_{\rm ours(Q)}$ denotes the second choice.

We first study $q_{\rm ours}$. We note that the result from \cite{liu2020thermodynamic} is only better than $q_{\rm ours}$ for small values of the transition probability ($P(A|B) < 0.2$). For the whole range, $q_{\rm LGU}$ can be $6$ orders of magnitude smaller than the quantity it is trying to lower bound. More importantly, $q_{\rm LGU}$ predicts the opposite trend for a large proportion of the transition probabilities. In sharp contrast, our proposed bound agrees in trend with the bound everywhere. Also, it is important to note that $q_{\rm ours}$ is tight for the two ends of the transition probabilities, while that of Ref.\cite{liu2020thermodynamic} is only tight to one side of $P(A|B)$ (i.e., only when $P(A|B)$ is small). For $q_{\rm ours(Q)}$, our bound is improved everywhere and performs similarly or better than $q_{\rm LGU}$ in lower bounding the fluctuation across all transition probabilities. This example shows that the freedom in choosing $Q$ can have strong physical implications and is useful in practice when one can take knowledge of the problem into consideration.

\section{Equilibrium Limit}
This section studies the equilibrium limit of the proposed relation (2). In particular, we show that it is a meaningful lower bound of any observable $f$ under consideration.

For simplicity, we assume $\gamma =1$. Note that this assumption should be valid when we are very close to equilibrium. When the system is in equilibrium, the probability of the reversed dynamics should be equal to the forward dynamics $P^*([x]^*)=P([x])$. Thus, when the system is only slightly away from equilibrium, there should exist a perturbatively small parameter $\alpha$ such that 
\begin{equation}
    P^*([x]^*) = P([x]) + \alpha h([x])  = P_{\alpha}([x]),
\end{equation}
for a function $h([x])$ such that $\sum_{[x]} h([x])=0$. Recall that our bound can be written as
\begin{align}
    \textrm{Var}[f] &\geq \frac{\left[\langle f\rangle - \langle f\rangle_{P^*([x]^*)}\right]^2}{\langle e^{-2\Delta S}\rangle -1}\\
    &= \frac{\left[\langle f\rangle_{P_{0}} - \langle f\rangle_{P_{\alpha}}\right]^2}{\left\langle \frac{P_\alpha^2}{P_0^2} \right\rangle -1}.
\end{align}
In the limit of zero $\alpha$, both the denominator and the numerator of the right-hand side becomes zero, whereas the ratio does not tend to zero in general and remains a meaningful lower bound of the variance of the observable consideration.

In fact, in the limit $\alpha \to 0^+$, we have
\begin{equation}
    \textrm{Var}[f] \geq \frac{\left( \frac{d}{d\alpha} \langle  f \rangle_{P_0} \right)^2}{\left\langle -\frac{\partial^2}{\partial \alpha^2} \log P_0 \right\rangle},
\end{equation}
which is a Cramer-Rao's bound when treating $\alpha$ as a parameter of the distribution $P([x])$, and is nontrivial in general. For example, $\frac{d}{d\alpha} \langle  f \rangle_{P_0}$ can be seen as the susceptibility of observable $f$ to an external perturbation controlled by $\alpha$, and this inequality can thus be seen as a form of the fluctuation-response theorems.

%\bibliography{ref}